# Tuning the Planarity of an Aromatic Thianthrene-Based Molecule on Au(111)

Kwan Ho Au-Yeung[1], Suchetana Sarkar[1], Sattwick Haldar[2], Pranjit Das[1], Tim Kühne[1], Dmitry A. Ryndyk[3,4,5], Preeti Bhauriyal[4], Stefan Kaskel[2], Thomas Heine[4], Gianaurelio Cuniberti[3], Andreas Schneemann[2], Francesca Moresco*[1]

[1]Center for Advancing Electronics Dresden, TU Dresden, 01062 Dresden, Germany

[2]Inorganic Chemistry I, TU Dresden, 01062 Dresden, Germany

[3]Institute for Materials Science, TU Dresden, 01062 Dresden, Germany

[4]Theoretical Chemistry, TU Dresden, 01062 Dresden, Germany

[5]Leibniz-Institute of Polymer Research, Hohe Straße 6, 01069 Dresden

ABSTRACT. Non-planar aromatic molecules are interesting systems for organic electronics and optoelectronics applications due to their high stability and electronic properties. By using scanning tunneling microscopy and spectroscopy, we investigated thianthrene-based molecules adsorbed on Au(111), which are non-planar in the gas phase and the bulk solid state. Varying the molecular coverage leads to the formation of two different kinds of self-assembled structures: close-packed islands and quasi one-dimensional chains. We found that the molecules are non-planar within the close-packed islands, while the configuration is planar in the molecular chain and for single adsorbed molecules. Using vertical tip manipulation to isolate a molecule from the island, we demonstrate the conversion of a non-planar molecule to its planar configuration. We discuss the two different geometries and their electronic properties with the support of density functional theory calculations.

INTRODUCTION

The planarity of polycyclic molecules has been studied for several decades due to its influence on the aromaticity of π-conjugated systems.[1-4] For non-planar molecules, the introduction of bond angle strain by, *e.g.*, heteroatoms or overcrowded structures,[5-6] causes geometric deformation of the chemical structure, and thus a bending of the π-system is observed. In some cases, such non-planar aromatic structures can preserve a conjugated π-system accompanied by some unique advantage, for instance conquering aggregation caused quenching, providing extraordinary stability and unique electronic properties for the applications in organic electronics and organic energy storage systems.[7-8]

Recently Haldar *et al.* reported the preparation of a two dimensional-layered covalent organic framework from 2,3,5,6-tetrafluoroterephthalonitrile and triphenylene-2,3,6,7,10,11-hexathiol, in which the organic building blocks are linked through dithiine bridges.[7] Interestingly, in contrast to structurally closely related compounds *i.e.*, dioxines[9] or phenazines,[10] the dithiine unit is not planar with a C-S-C dihedral angle of 101°. In parallel to the dithiine linked framework structure, also a model compound - benzo[5,6][1,4]dithiino[2,3-b]thianthrene-6,13-dicarbonitrile (bTEpCN) - featuring two dithiine units was isolated from the reaction of 2,3,5,6-tetrafluoroterephthalonitrile and benzene-1,2-dithiol, and the crystal structure could be solved with an angle of 120° between neighboring benzene rings (see Figure 1).

Another group of researchers solved the single crystal structure in parallel to us.[11] In their work the molecule served as emitters for thermally activated delayed fluorescence and room temperature phosphorescence with potential application in organic light emitting diodes, highlighting its interesting optoelectronic properties. During our structural investigation of the dithiine linked covalent organic framework (COF), theoretical calculations of the molecules in gas phase using natural bond orbital analysis suggested that the conjugation within the dithiine is not necessarily fully broken by the non-planarity of the aromatic rings neighboring the dithiine bridge. One of the sulfur lone pairs is hybridized featuring s and p character, while the other is only featuring p character. Interestingly, the dithiine core shows weak overlap of the sulfur's nonbonding electrons with the 2p orbital of the neighboring carbon. A 16π electron dithiine unit is generated through this sp mixing, leading to a rare case of nonplanar weak aromaticity. Furthermore, it is known that dithiine units can be reduced. The loss of a single electron leads to a dithiine radical, which has a fully planar structure.[12-13] All in all, the nonplanarity of thianthrene-based systems at the single molecule level still remains challenging to access.

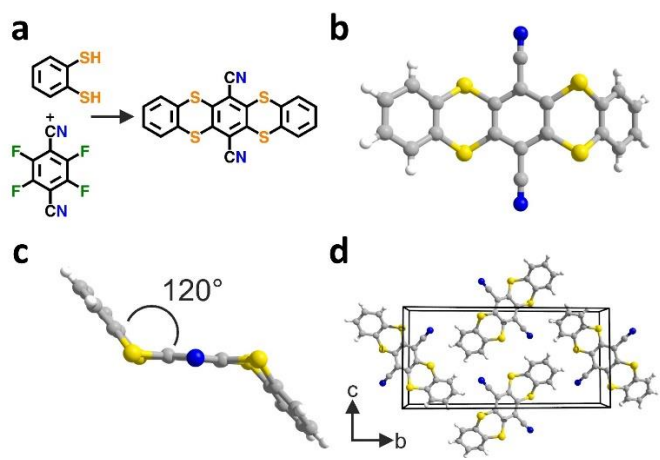

**Figure 1.** bTEpCN compound in single crystal structure. (a) Schematic depiction of the synthesis of the bTEpCN compound. (b) Top view of a single bTEpCN molecule from the single crystal structure. (c) Side view of the bTEpCN molecule, highlighting the non-planarity in the crystal structure. (d) Packing of the bTEpCN molecules in the crystal structure. C, H, N and S are depicted in grey, white, blue and yellow, respectively.

The described properties of thianthrene-based systems motivated us to investigate the planarity of bTEpCN adsorbed on the Au(111) surface. Experiments have been performed by scanning tunneling microscopy (STM) and spectroscopy (STS) operated at low temperature (5 K) under ultra-high vacuum (UHV) conditions, supported by density functional theory (DFT) calculations. To probe the molecular geometric deformation, the high spatial resolution in real space provided by STM offers the possibility to investigate the planarity of molecules at the single-molecule level.[14-20] We employ STM manipulation to convert the molecules from non-planar to planar configuration, revealing that the intermolecular interactions within the molecular islands stabilize the non-planar configuration. We compare the electronic structures by using STS measurements.

METHODS

The synthesis and characterization of benzo[5,6][1,4]dithiino[2,3-b]thianthrene-6,13-dicarbonitrile is described in the Supporting Information file.

bTEpCN molecules were sublimated from a quartz crucible ($T_{evap} \approx 240\ °C$) on the clean Au(111) surface kept at room temperature under ultrahigh vacuum (UHV) conditions. Before evaporation, the Au(111) single crystal was cleaned by subsequent cycles of $Ar^+$ sputtering and annealing to 450 °C. STM experiments were performed using a custom-built instrument operating at a low temperature of T = 5 K under ultrahigh vacuum ($p \approx 1 \times 10^{-10}$ mbar).

STM images were taken in constant current mode. CO molecules were deposited with a very low surface coverage onto the cold sample and then deliberately picked up by the tip of the STM to functionalize the apex. High resolution STM images were recorded in constant height mode with the CO-functionalized tip apex. A CO tip enhances the resolution of a molecule as compared to a bare tip. This is because for the usual s-wave tip states, STM images resemble the local densities of state, given by the modulus squared of sample wave functions at the Fermi energy.[21] However, upon functionalizing with CO, the tip states are changed to p-wave,[22] which show the spatial derivatives of the sample wave functions.[23] This technique is therefore used to resolve the internal structure of the molecule, which would otherwise be impossible with a metal tip.

STS dI/dV spectra and maps were measured using lock-in detection with a modulation frequency of 833 Hz.

All vertical manipulation for controllably picking up the molecules were performed in constant-height mode. The pick-up manipulation involves: (1) allowing the tip to vertically approach the molecules under a small bias (e.g., V = 10 mV), (2) vertically driving the tip near to the molecule non-destructively, (3) then retracting the tip to normal scanning position, and (4) eventually re-depositing the molecule on the tip apex by a similar procedure of pick-up manipulation far away from the area of interest. When necessary, tip was re-shaped by gently crashing on the gold surface far away from the area of interest.

For geometry optimization, we used the DFT method as implemented in the CP2K software package[30] (cp2k.org) with the Quickstep module.[24] The Perdew-Burke-Ernzerhof exchange-correlation functional,[25] the Goedecker-Teter-Hutter pseudo-potentials[26] and the valence double-$\zeta$ basis sets, in combination with the DFT-D2 method of Grimme[27] for van der Waals (vdW) correction were applied. We used 6 layers of gold, where the 3 upper layers were allowed to be relaxed (planar supercell 29.8 x 19.9 Å, vacuum size 40 Å, maximum force 4.5 x $10^{-5}$ a.u.). The calculations of STM topography images and dI/dV images were performed by the DFTB+XT code from TraNaS OpenSuite (tranas.org/opensuite), partially based on the DFTB+[28-29] software package. We also used the density functional based tight-binding method with auorg-1-1 parametrization[30-31] as implemented in the DFTB+ package. We considered a realistic atomistic system including the STM tip and the substrate, both connected to semi-infinite electrodes. The simulation of STM images in the constant-current mode was done based on the current calculations by the Green function technique[32]. The calculated data was analyzed, and the images were generated by the PyMOL Molecular Graphics System, Version 2.4 open-source build, Schrödinger, LLC.

RESULTS AND DISCUSSION

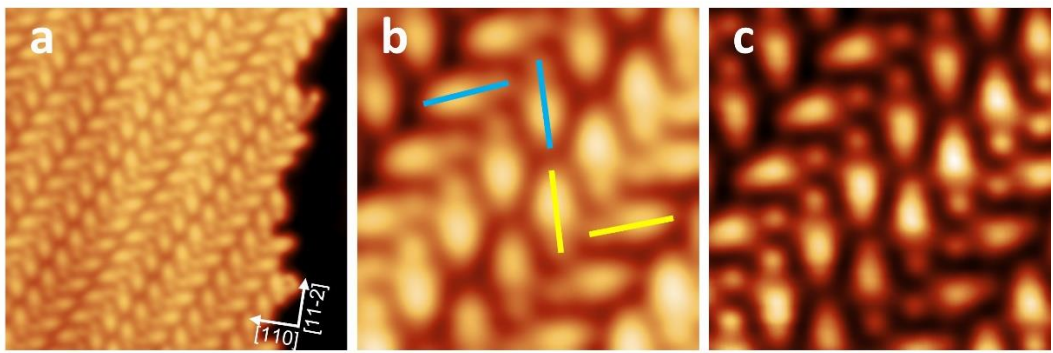

**Figure 2.** Self-assembled molecular islands consisting of bTEpCN molecules adsorbed on Au(111). STM images (a) 20 nm x 20 nm and (b) 5 nm x 5 nm were taken at V = 0.2 V and I = 20 pA. (c) Constant height STM image (5 nm x 5 nm) with a CO-functionalized tip taken at V = 12 mV at the same position of (b). The blue and yellow stripes in (b) reveal the two observed arrangement directions.

After the thermal deposition ($T_{evap}$ = 234 °C) of benzo[5,6][1,4]dithiino[2,3-b]thianthrene-6,13-dicarbonitrile (bTEpCN) molecules under UHV conditions on a clean Au(111) surface kept at room temperature, STM experiments were performed after cooling the sample to T = 5 K.

An overview STM image (Figure 2a) shows ordered self-assembled molecular islands growing from the step edges of the Au(111) surface at a moderate coverage. The stripes reveal two arrangement directions as indicated in Figure 2b by blue (up) and yellow (down) lines respectively, where they have no preferred direction (~50%). Within the islands, the molecules appear in the STM images of Figure 2b as a pear shape, showing a smaller protrusion at one end and a larger protrusion at the center, suggesting a non-planar asymmetric conformation. With a CO-functionalized STM tip providing high-resolution images in constant height mode (see Methods for details), Figure 2c shows that one of the peripheral benzene rings appears more intense than the other end. The angle between two molecules within a molecular row is about 81°, and the distance between the molecules in the same alignment measured from center-to-center is about 1.1 nm, very similar to the molecule-molecule distance in the single crystal structure.[11] DFT calculations of a close-packed molecular island have a very good agreement with the experimental results in terms of the geometry and the intermolecular distances to the neighbors, confirming that the molecules are non-planar (Figure S9). As we will discuss in detail by the manipulation experiments below, the non-planarity is reduced with respect to the single crystal structure (Figure 1c) due to the van der Waals interaction with the gold surface.

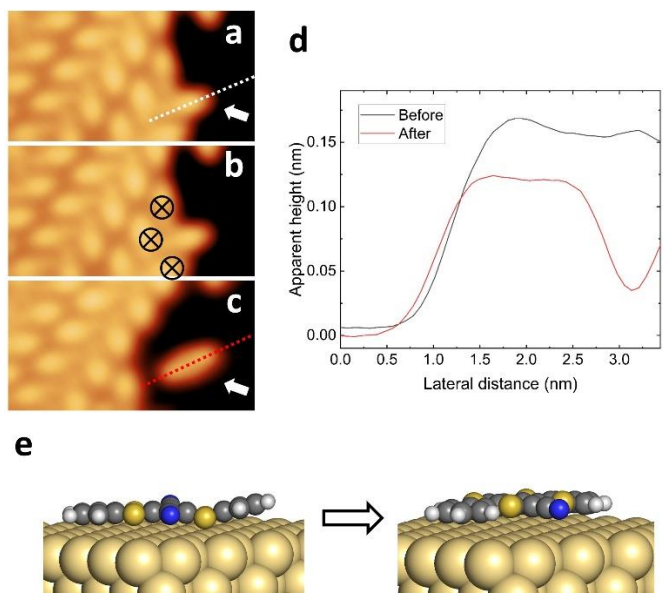

**Figure 3.** Indirect isolation for converting the nonplanarity of a bTEpCN molecule. (a) The target non-planar molecule located at the edge of the island is marked. (b) Picking up (V = 10 mV, I = 5 – 8 nA) the three surrounding neighbors (marked with "x") as an indirect isolation. (c) The remaining target molecule becomes planar. All STM images (7 nm x 3.7 nm) were taken under V = 0.2 V and I = 20 pA. (d) Line profiles show the different appearances comparing before and after the planarization. (e) DFT calculated adsorption geometries (side view) of a non-planar and a planarized bTEpCN molecule on Au(111).

Numbers of examples on conformational switching of molecules on surface have been shown, where it is possible to convert molecular conformations by means of STM tip manipulation.[20, 33-36] However, while STM voltage pulses do not yield any conformational changes in the present study, reducing the quantity of neighbor molecules revealed that the intermolecular interaction plays a major role in the adsorption geometry, shedding light on the origin of non-planarity on surface.

Figure 3 shows a STM manipulation sequence of removing the neighbors that directly interact with the target molecule (arrow) at the edge of the island (neighbors = 3). Indirect isolation of molecules within the islands was done by vertical manipulation, picking up the neighbor molecules by the STM apex and depositing back onto the surface at another location far away (> 100 nm) from the area of interest (not shown). By controllably picking up (removing) the neighbors, we can avoid any undesired tip-induced excitations or direct change of conformations on the target molecule, also avoiding any damage to the gold surface. When necessary, the STM tip is re-shaped by gently crashing on the gold surface far away from the area of interest. Interestingly, removing the first two neighbors does not have any effect on the adsorption geometry, while after removing the third neighbor (Figure 3b), the target molecule appears differently (Figure 3c), revealing a symmetric and planar topographic appearance. Further analysis on the line profiles (Figure 3d) confirms that the apparent height of the target molecule has been changed after indirect isolation.

Another strategy that leads to the same result where the manipulation procedure was carried out in the middle of the island (Figure S7). In this case, we stepwise removed the six neighbors individually, and found that the molecule requires at least three neighbors to maintain its non-planar conformation. No specific sequential order (*e.g.,* which molecular neighbor to be picked up at last) during the removal procedure for planarization was observed. We do not observe any differences between the non-planar molecules at the edge of the island and those within the island (with all six neighbors). Moreover, we have not observed any isolated stand-alone non-planar molecules. The non-planarity is also irreversible, possibly due to the fact that it is not possible to bring new neighbors to an isolated molecule.

To gain more information about the two different adsorption geometries, we performed DFT calculations to simulate STM images and identify the adsorption geometries of the molecules. Figure 3e shows the side view of the non-planar and planarized molecules on the Au(111) surface. Note that at the top picture, only one molecule from the island is shown for clarity, where the non-planar molecules only exist inside dense-packed islands, and the non-planarity is supported by this supramolecular structure. The calculations show that the C-S-C bending happens on one end of the molecule while the other end remains flat. The calculated C-S-C dihedral angle is 147° on one side, and the other side is about 180°, confirming that the non-planarity is reduced on the Au(111) surface respect to the gas phase (Figure 1c) due to the van der Waals interaction with the gold surface. On the other hand, the isolated stand-alone planar molecule is symmetrically flat and has a C-S-C dihedral angle of about 180° on both sides.

The above results demonstrate that the intermolecular interactions (*i.e.* a combination of van der Waals forces and Coulomb interaction of partial atomic charges screened by a surface electron) play an important role in stabilizing the non-planarity. In other words, this may suggest that with lower molecular coverage, where the molecules are no longer assembled as close-packed islands can lead to different self-assembled structures and conformations.

Indeed, we observed at lower local coverage another type of self-assembled structure (Figure 4), which are quasi one-dimensional molecular chains along the fcc regions of the Au(111) surface. The molecular chains tend to avoid the solitons from the Au(111) reconstruction and re-arrange to a more energetically favorable site within the fcc regions, as shown in Figure 4a. In a few cases (Figure S8) a coexistence of islands and chains is observed. A close-up STM image (Figure 4b) reveals the angle and inter-molecular distance between the neighbors with 72° and 0.96 nm, respectively. The calculated adsorption geometries of the quasi-one-dimensional chain are in very good agreement with experimental results, giving the angle and the inter-molecular distance between the neighbors with 70° and 1 nm, respectively (Figure S9). More interestingly, the molecules appear very similarly (Figure 4b) to the isolated planar molecule in Figure 3.

After annealing the surface at 150° C, both islands of non-planar molecules and lines of planar ones are still present on the surface, indicating that both forms are stable until that temperature. However, after further annealing to 200°C we observe only chains of planar molecules, while at 220° both forms are desorbed. These results confirm that the planar version is energetically more stable than the non-planar one, in agreement with the DFT calculations.

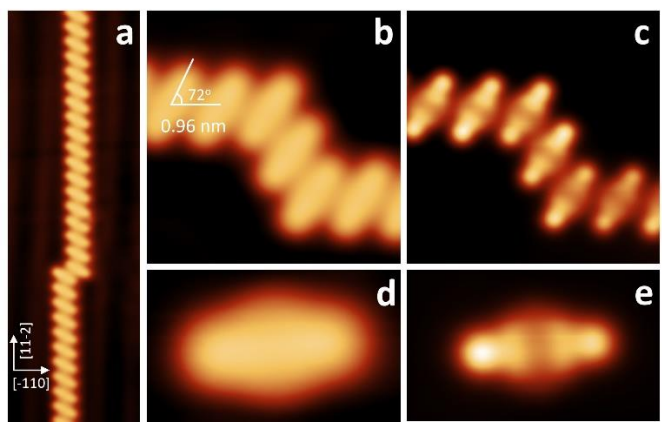

**Figure 4.** Self-assembled molecular chains and isolated molecule. (a) STM image (8 nm x 25 nm) of a molecular chain along the fcc region of the Au(111) surface. (b) Close-up of a molecular chain (5 nm x 5 nm). (d) An isolated bTEpCN molecule (2.5 nm x 1.5 nm). (c, e) Corresponding high-resolution STM images with same sizes taken at V = 12 mV. Constant current STM images were taken under (a) V = 0.5 V, I = 20 pA, (b, d) V = 0.2 V, I = 20 pA.

To compare the two molecular conformations, we isolated a single molecule from a molecular chain by removing two neighbor molecules using the same procedure demonstrated before (Figure 4d), where the isolated molecule shows the same appearance as in Figure 3c. We then investigate the molecular structure with high-resolution STM images with a CO-functionalized tip (Figure 4c and 4e). The isolated molecule from Figure 3c has the same adsorption geometry as the molecules in the molecular chains (Figure 4b) and the isolated molecule from the chain (Figure 4d). In other words, we found that the same molecular species adsorbs in different ways in the two different self-assembled structures on the same sample, depending on the coverage-dependent intermolecular interactions.

To further investigate the bTEpCN molecules in their planar and non-planar configurations, we studied the electronic structure of the molecules by STS (Figures 5 and 6). In Figure 5a we first plot the dI/dV spectrum of an isolated bTEpCN planar molecule after removing the neighbors by vertical manipulation. Because the contrast in the dI/dV spectra is relatively weak especially at low energies, we assign the electronic resonance values by sweeping the bias voltage stepwise and recording the differential conductance maps (See the complete sweep in Figure S10 – S13).

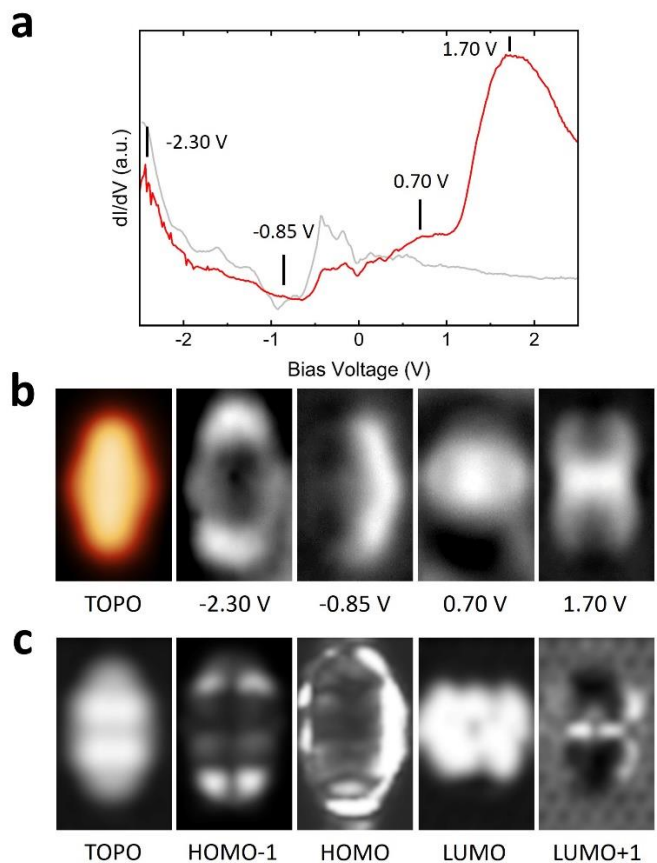

**Figure 5.** dI/dV spectra and maps of an isolated planar bTEpCN molecule. (a) dI/dV spectrum of the planar (red line) configuration, in grey for comparison the dI/dV spectrum recorded on the clean Au(111). The spectra are averaged over 10 curves The position of the electronic resonances is indicated, as determined by the contrast maxima in the dI/dV maps shown in (b). (b) STM topography and the dI/dV maps corresponding to the electronic resonances taken with I = 300 pA

(1.5 nm x 2.5 nm) (c) Corresponding DFT calculated STM topographic image and molecular orbitals.

For the isolated planar molecule, the dI/dV spectrum of Figure 5a shows three visible peaks within the range accessible by STS (-2.5 V < V < 2.5 V). Below the Fermi energy, we observe a resonance at V = -2.30 V. By recording the dI/dV maps sweeping the bias voltage, we identify the differential conductance map at this energy, and find a further feature with an enhanced contrast at V =-0.85 V (Figure 5b and the complete sweep in Figure S12-S13). Above the Fermi level we observe peaks with corresponding differential conductance maps at V = 0.70 V and V = 1.70 V. In principle, the mono-electronic approximation is not exactly valid while tunnelling through a molecule. Therefore, several molecular orbitals are needed to describe electronic tunneling through a molecule and the electronic resonances observed in STS.[37] Normally however at least the frontier resonances are mainly determined by the highest occupied molecular orbital (HOMO) and the lowest unoccupied molecular orbital (LUMO) respectively.

We performed DFT-based calculations of STM topography images and dI/dV maps corresponding to the resonant molecular orbitals of the isolated planar molecule on Au(111) (Figure 5c). The good agreement between experiment and calculations confirms the assigned conformation of the isolated molecule on Au(111), also indicating that the observed resonances are well described by the molecular orbitals of the adsorbed molecule HOMO-1 (V = -2.30 V), HOMO (V =-0.85 V), LUMO (V = 0.70 V) and LUMO+1 (V = 1.70 V) respectively.

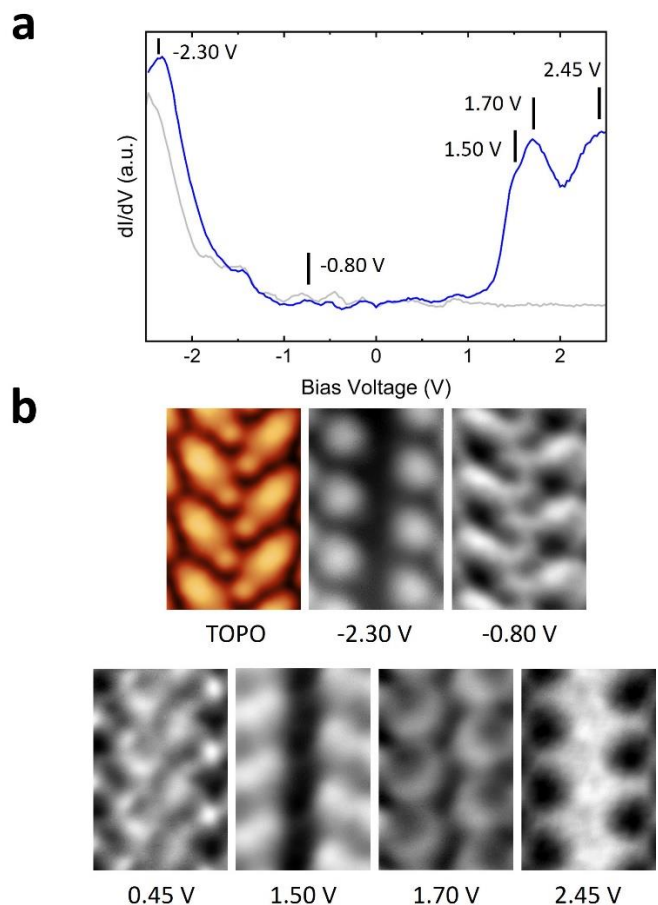

**Figure 6.** dI/dV spectra and maps of non-planar bTEpCN molecules assembled in the island non-planar conformation. (a) dI/dV spectrum of the non-planar (blue line) configuration, in grey for comparison the dI/dV spectrum on the clean Au(111). (b) STM topography and the corresponding dI/dV maps taken with I = 300 pA (2.3 nm x 3.5 nm).

In Figure 6a, the dI/dV spectrum of a non-planar molecule in an island shows four visible peaks within the same range. Similar to the case of a single planar molecule, we observe below the Fermi energy an intense peak at V = -2.30 V. When we stepwise sweep the bias values recording the dI/dV maps (Figure 6b and S10 – S11), we identify also in this case an enhanced contrast at V = -0.8 V. We can conclude that the filled-states resonances of the non-planar molecules in the island are in good agreement with the planar single molecule. Furthermore, in the dI/dV maps at negative bias voltage one can recognize the contrast of the single molecules.

Above the Fermi energy, the dI/dV spectrum of the non-planar molecule forming an island shows three resonances at 1.50 V, 1.70 V, and 2.45 V respectively, which are also well visible in the dI/dV maps. Furthermore, we identify from the dI/dV map sweep an emerging relative intense feature at the sulfur moieties of the non-planar molecules at 0.45 V, which is not observed for the isolated planar molecule. It is however important to note that the comparison between planar and non-planar case is complicated by the close interaction between neighbor molecules in the island for the non-planar case, which can influence the observed electronic structure. Also for this reason the comparison with calculated orbitals is in this case less straightforward (see Figure S14).

CONCLUSIONS

In this study, we have investigated the non-planar aromatic bTEpCN molecules adsorbed on the Au(111) surface. Two different self-assembled structures are identified after the molecular adsorption, resulting in a planar and a non-planar configuration on the surface, depending on the intermolecular interactions. Single non-planar molecules within the molecular island can be converted to the planar configuration after reducing the number of neighbors by STM vertical manipulation. Our results also indicate a slight influence of the conformation of the molecules on the position and localization of the electronic resonances at positive voltages, contributing to understand the relation between planarity and aromaticity of polycyclic molecular systems.

ASSOCIATED CONTENT

**Supporting Information.** The following files are available free of charge:

- A supporting information PDF file is containing the synthesis and chemical characterization of the molecule and further experimental results and supplementary calculations.

AUTHOR INFORMATION

**Corresponding Authors.** *Email: francesca.moresco@tu-dresden.de

**Author Contributions.** The manuscript was written through contributions of all authors. All authors have given approval to the final version of the manuscript.

ACKNOWLEDGMENT

This work was funded by the German Research Foundation (DFG) by the Collaborative Research Centre (CRC) 1415. Support from the European Innovation Council (EIC) under the project ESiM (grant agreement No 101046364), and from the DFG Project 43234550 is acknowledged.

S.H. thanks the DFG Priority Programme "Polymer-based Batteries" (SPP 2248) for the financial support.

A.S. gratefully acknowledges the Fonds der chemischen Industrie for a Liebig Fellowship. We thank the Center for Information Services and High Performance Computing (ZIH) at TU Dresden for computational resources.

TOC GRAPHIC

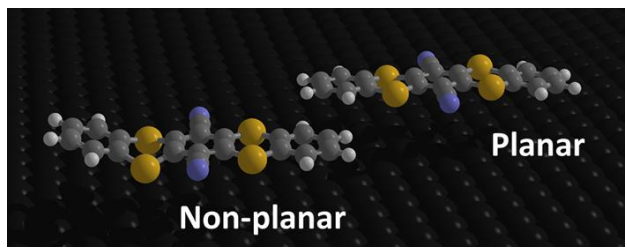

REFERENCES

1. Antić, M.; Furtula, B.; Radenković, S. Aromaticity of Nonplanar Fully Benzenoid Hydrocarbons. *J. Phys. Chem. A* **2017,** *121*, 3616-3626.
2. Bodwell, G. J.; Bridson, J. N.; Cyrañski, M. K.; Kennedy, J. W. J.; Krygowski, T. M.; Mannion, M. R.; Miller, D. O. Nonplanar Aromatic Compounds. 8.1 Synthesis, Crystal Structures, and Aromaticity Investigations of the 1,N-Dioxa[N](2,7)Pyrenophanes. How Does Bending Affect the Cyclic Π-Electron Delocalization of the Pyrene System? *J. Org. Chem.* **2003,** *68*, 2089-2098.
3. Li, Y.; Yagi, A.; Itami, K. Synthesis of Highly Twisted, Nonplanar Aromatic Macrocycles Enabled by an Axially Chiral 4,5-Diphenylphenanthrene Building Block. *J. Am. Chem. Soc.* **2020,** *142*, 3246-3253.
4. Nestoros, E.; Stuparu, M. C. Corannulene: A Molecular Bowl of Carbon with Multifaceted Properties and Diverse Applications. *Chem. Commun.* **2018,** *54*, 6503-6519.
5. Kawahara, K. P.; Ito, H.; Itami, K. One-Step Synthesis of Polycyclic Thianthrenes from Unfunctionalized Aromatics by Thia-Apex Reactions. *Org. Chem. Front.* **2023,** *10*, 1880-1889.
6. Saha, P. K.; Mallick, A.; Turley, A. T.; Bismillah, A. N.; Danos, A.; Monkman, A. P.; Avestro, A.-J.; Yufit, D. S.; McGonigal, P. R. Rupturing Aromaticity by Periphery Overcrowding. *Nat. Chem.* **2023,** *15*, 516-525.
7. Haldar, S.; Wang, M.; Bhauriyal, P.; Hazra, A.; Khan, A. H.; Bon, V.; Isaacs, M. A.; De, A.; Shupletsov, L.; Boenke, T. *et al.* Porous Dithiine-Linked Covalent Organic Framework as a Dynamic Platform for Covalent Polysulfide Anchoring in Lithium–Sulfur Battery Cathodes. *J. Am. Chem. Soc.* **2022,** *144*, 9101-9112.
8. Speer, M. E.; Kolek, M.; Jassoy, J. J.; Heine, J.; Winter, M.; Bieker, P. M.; Esser, B. Thianthrene-Functionalized Polynorbornenes as High-Voltage Materials for Organic Cathode-Based Dual-Ion Batteries. *Chem. Commun.* **2015,** *51*, 15261-15264.
9. Zhang, B.; Wei, M.; Mao, H.; Pei, X.; Alshmimri, S. A.; Reimer, J. A.; Yaghi, O. M. Crystalline Dioxin-Linked Covalent Organic Frameworks from Irreversible Reactions. *J. Am. Chem. Soc.* **2018,** *140*, 12715-12719.
10. Guo, J.; Xu, Y.; Jin, S.; Chen, L.; Kaji, T.; Honsho, Y.; Addicoat, M. A.; Kim, J.; Saeki, A.; Ihee, H. *et al.* Conjugated Organic Framework with Three-Dimensionally Ordered Stable Structure and Delocalized Π Clouds. *Nat. Commun.* **2013,** *4*, 2736.

11. Li, M.; Xie, W.; Cai, X.; Peng, X.; Liu, K.; Gu, Q.; Zhou, J.; Qiu, W.; Chen, Z.; Gan, Y. *et al.* Molecular Engineering of Sulfur-Bridged Polycyclic Emitters Towards Tunable Tadf and Rtp Electroluminescence. *Angew. Chem. Int. Ed.* **2022,** *61*, e202209343.

12. Beck, J.; Bredow, T.; Tjahjanto, R. T. Thianthrene Radical Cation Hexafluorophosphate. *Z. Naturforsch. B* **2009,** *64*, 145-152.

13. Wang, S.; Yuan, J.; Xie, J.; Lu, Z.; Jiang, L.; Mu, Y.; Huo, Y.; Tsuchido, Y.; Zhu, K. Sulphur-Embedded Hydrocarbon Belts: Synthesis, Structure and Redox Chemistry of Cyclothianthrenes. *Angew. Chem. Int. Ed.* **2021,** *60*, 18443-18447.

14. Wu, F.; Barragán, A.; Gallardo, A.; Yang, L.; Biswas, K.; Écija, D.; Mendieta-Moreno, J. I.; Urgel, J. I.; Ma, J.; Feng, X. Structural Expansion of Cyclohepta[Def]Fluorene Towards Azulene-Embedded Non-Benzenoid Nanographenes. *Chem. Eur. J.* **2023,** *29*, e202301739.

15. Chen, P.; Joshi, Y. V.; Metz, J. N.; Yao, N.; Zhang, Y. Conformational Analysis of Nonplanar Archipelago Structures on a Cu (111) Surface by Molecular Imaging. *Energy Fuels* **2020,** *34*, 12135-12141.

16. Urgel, J. I.; Di Giovannantonio, M.; Eimre, K.; Lohr, T. G.; Liu, J.; Mishra, S.; Sun, Q.; Kinikar, A.; Widmer, R.; Stolz, S. *et al.* On-Surface Synthesis of Cumulene-Containing Polymers Via Two-Step Dehalogenative Homocoupling of Dibromomethylene-Functionalized Tribenzoazulene. *Angew. Chem. Int. Ed.* **2020,** *59*, 13281-13287.

17. Xu, K.; Urgel, J. I.; Eimre, K.; Di Giovannantonio, M.; Keerthi, A.; Komber, H.; Wang, S.; Narita, A.; Berger, R.; Ruffieux, P. *et al.* On-Surface Synthesis of a Nonplanar Porous Nanographene. *J. Am. Chem. Soc.* **2019,** *141*, 7726-7730.

18. Ammon, M.; Sander, T.; Maier, S. On-Surface Synthesis of Porous Carbon Nanoribbons from Polymer Chains. *J. Am. Chem. Soc.* **2017,** *139*, 12976-12984.

19. Fiedler, B.; Rojo-Wiechel, E.; Klassen, J.; Simon, J.; Beck, J.; Sokolowski, M. Ordered Structures of Two Sulfur Containing Donor Molecules on the Au(111) Surface. *Surf. Sci.* **2012,** *606*, 1855-1863.

20. Pavliček, N.; Fleury, B.; Neu, M.; Niedenführ, J.; Herranz-Lancho, C.; Ruben, M.; Repp, J. Atomic Force Microscopy Reveals Bistable Configurations of Dibenzo[a,H]Thianthrene and Their Interconversion Pathway. *Phys. Rev. Lett.* **2012,** *108*, 086101.

21. Tersoff, J.; Hamann, D. R. Theory of the Scanning Tunneling Microscope. *Phys. Rev. B* **1985,** *31*, 805-813.

22. Gross, L.; Moll, N.; Mohn, F.; Curioni, A.; Meyer, G.; Hanke, F.; Persson, M. High-Resolution Molecular Orbital Imaging Using a $P$-Wave Stm Tip. *Phys. Rev. Lett.* **2011,** *107*, 086101.

23. Chen, C. J., *Introduction to Scanning Tunneling Microscopy* Oxford University Press: New York, 1993.

24. VandeVondele, J.; Krack, M.; Mohamed, F.; Parrinello, M.; Chassaing, T.; Hutter, J. Quickstep: Fast and Accurate Density Functional Calculations Using a Mixed Gaussian and Plane Waves Approach. *Comput. Phys. Commun.* **2005,** *167*, 103-128.

25. Perdew, J. P.; Burke, K.; Ernzerhof, M. Generalized Gradient Approximation Made Simple. *Phys. Rev. Lett.* **1996,** *77*, 3865-3868.

26. Goedecker, S.; Teter, M.; Hutter, J. Separable Dual-Space Gaussian Pseudopotentials. *Phys. Rev. B* **1996,** *54*, 1703-1710.

27. Grimme, S.; Antony, J.; Ehrlich, S.; Krieg, H. A Consistent and Accurate Ab Initio Parametrization of Density Functional Dispersion Correction (Dft-D) for the 94 Elements H-Pu. *J. Chem. Phys.* **2010,** *132*, 154104.

28. Hourahine, B.; Aradi, B.; Blum, V.; Bonafé, F.; Buccheri, A.; Camacho, C.; Cevallos, C.; Deshaye, M. Y.; Dumitrică, T.; Dominguez, A. *et al.* Dftb+, a Software Package for Efficient Approximate Density Functional Theory Based Atomistic Simulations. *J. Chem. Phys.* **2020,** *152*, 124101.
29. Pecchia, A.; Penazzi, G.; Salvucci, L.; Di Carlo, A. Non-Equilibrium Green's Functions in Density Functional Tight Binding: Method and Applications. *New J. Phys.* **2008,** *10*, 065022.
30. Elstner, M.; Porezag, D.; Jungnickel, G.; Elsner, J.; Haugk, M.; Frauenheim, T.; Suhai, S.; Seifert, G. Self-Consistent-Charge Density-Functional Tight-Binding Method for Simulations of Complex Materials Properties. *Phys. Rev. B* **1998,** *58*, 7260-7268.
31. Fihey, A.; Hettich, C.; Touzeau, J.; Maurel, F.; Perrier, A.; Köhler, C.; Aradi, B.; Frauenheim, T. Scc-Dftb Parameters for Simulating Hybrid Gold-Thiolates Compounds. *J. Comput. Chem.* **2015,** *36*, 2075-2087.
32. Ryndyk, D. A., *Theory of Quantum Transport at Nanoscale*. Springer International Publishing: 2016; Vol. 184.
33. Moresco, F.; Meyer, G.; Rieder, K.-H.; Tang, H.; Gourdon, A.; Joachim, C. Conformational Changes of Single Molecules Induced by Scanning Tunneling Microscopy Manipulation: A Route to Molecular Switching. *Phys. Rev. Lett.* **2001,** *86*, 672-675.
34. Sarkar, S.; Au-Yeung, K. H.; Kühne, T.; Waentig, A.; Ryndyk, D. A.; Heine, T.; Cuniberti, G.; Feng, X.; Moresco, F. Adsorption and Reversible Conformational Change of a Thiophene Based Molecule on Au(111). *Sci. Rep.* **2023,** *13*, 10627.
35. Qi, J.; Gao, Y.; Jia, H.; Richter, M.; Huang, L.; Cao, Y.; Yang, H.; Zheng, Q.; Berger, R.; Liu, J. *et al.* Force-Activated Isomerization of a Single Molecule. *J. Am. Chem. Soc.* **2020,** *142*, 10673-10680.
36. Soe, W.-H.; Shirai, Y.; Durand, C.; Yonamine, Y.; Minami, K.; Bouju, X.; Kolmer, M.; Ariga, K.; Joachim, C.; Nakanishi, W. Conformation Manipulation and Motion of a Double Paddle Molecule on an Au(111) Surface. *ACS Nano* **2017,** *11*, 10357-10365.
37. Portais, M.; Joachim, C. Hole–Electron Quantum Tunnelling Interferences through a Molecular Junction. *Chem. Phys. Lett.* **2014,** *592*, 272-276.